\documentclass[aps,twocolumn,superscriptaddress]{revtex4-2}

\usepackage{amsmath,amssymb,amsfonts,color,graphicx,tabularx}

\usepackage[unicode=true,colorlinks=true]{hyperref}

\hypersetup{linkcolor=blue,citecolor=blue,urlcolor=blue}

\begin{document}

\title{Quantum phases of two-component bosons on the Harper-Hofstadter ladder}

\author{Jian-Dong Chen}
\affiliation{Department of Physics, Harbin Institute of Technology, Harbin 150001, China}
\affiliation{Shenzhen Institute for Quantum Science and Engineering and Department of Physics, Southern University of Science and Technology, Shenzhen 518055, China}

\author{Hong-Hao Tu}
\affiliation{Institut f\"ur Theoretische Physik, Technische Universit\"at Dresden, 01062 Dresden, Germany}

\author{Ying-Hai Wu}
\email{yinghaiwu88@hust.edu.cn}
\affiliation{School of Physics and Wuhan National High Magnetic Field Center, Huazhong University of Science and Technology, Wuhan 430074, China}

\author{Zhi-Fang Xu}
\email{xuzf@sustech.edu.cn}
\affiliation{Shenzhen Institute for Quantum Science and Engineering and Department of Physics, Southern University of Science and Technology, Shenzhen 518055, China}

\begin{abstract}
We study two-component bosons on the Harper-Hofstadter model with two legs. The synthetic magnetic fields for the two types of bosons point to either the same direction or opposite directions. The bosons have hardcore intra-species interaction such that there can be no more than one boson of the same type on each lattice site. For certain filling factors in the absence of inter-species interaction, each component realizes a vortex Mott insulator with rung current or a Meissner superfluid without rung current. The system undergoes phase transitions to other phases as inter-species interaction is turned on, which are characterized numerically using the density matrix renormalization group method and supplemented with analytical studies when possible. The vortex Mott insulator transits to a gapped Meissner phase without rung current and the Meissner superfluid transits to a gapped vortex phase with rung current. In both cases, we observe gapped spin density wave states that break certain ${\mathbb Z}_{2}$ symmetries.
\end{abstract}

\maketitle

\section{Introduction}
\label{intro}

In the past few years, great efforts have been devoted to create synthetic gauge fields in artificial quantum systems such as cold atoms, photonic crystals, and superconducting circuits~\cite{Dalibard2011,Goldman2016,Ozawa2019}. This is largely motivated by the interest on topological phases as many of them require gauge fields to be realized. The classical examples are quantum Hall states observed in the 1980s~\cite{Klitzing1980,Tsui1982}, where a strong magnetic field is applied to quench the kinetic energy and produce Landau levels for two-dimensional electron gases. The rise of topological insulators further boosts the interest in the physics community~\cite{Hasan2010,QiXL2011-1}. One fundamental insight in this discovery is that symmetries are needed to protect certain topologically non-trivial phenomena. If the protecting symmetries are not respected in a system, topological states may be destroyed without gap closing and reopening. One pivotal element for realizing topological insulators is spin-orbit couplings, which can be interpreted as non-Abelian gauge fields at the theoretical level. 

One model that has drawn widespread attention is the Harper-Hofstadter model~\cite{Harper1955,Hofstadter1976,Streda1982,Hatsugai1993,Bernevig-Book}. It was originally proposed in the context of solid state systems to understand charged particles moving in the presence of both magnetic field and periodic potential. The tight binding approximation is adopted to address the periodic potential. If a particle executes circular motion in a magnetic field, it would pick up a phase that reflects the magnetic flux enclosed by its trajectory, which is accounted for by complex hopping phases between lattice sites. The phases destroy the original lattice translational symmetry but magnetic translational symmetry is preserved if the flux per plaquette is a rational number. In such cases, the Bloch theorem is still applicable on the scale of magnetic unit cells and energy bands can be defined as usual. The Chern numbers of the energy bands are related to the flux per plaquette and the Hall conductance via the Diophantine equation. This model was not very useful from an experimental perspective for a long time~\cite{Weiss1990,Gerhardts1991,Albrecht1998,Nakamura1998} because the magnetic field applied to solid state systems is not strong enough to make lattice effects sufficiently important, as quantified by the ratio between magnetic length $\sqrt{hc/(eB)}$ and lattice constants. This challenge has been overcome in van der Waals heterostructures~\cite{Ponomarenko2013,Dean2013,Hunt2013}, where lattice periodicity is substantially altered due to the formation of Moire superlattice pattern. 

The pursuit of the Harper-Hofstadter model in artificial quantum systems has also been very furitful~\cite{Aidelsburger2013,Miyake2013,Mancini2015,Stuhl2015,Aidelsburger2015,Mittal2016,TaiEM2017,Roushan2017}. While most experiments are still limited in one way or another, the advances along this direction are impressive and many more interesting phenomena can be expected. The foremost difficulty in such systems is the absence of particles that carry electric charges and couple to the magnetic field. To this end, synthetic gauge fields that mimic the effect of actual magnetic fields but act on charge neutral particles have to be designed. For cold atoms in optical lattices, the complex hopping phases in the model can be achieved using laser-assisted tunneling, but this also results in considerable heating that impairs the stability of the system. In spite of such technical obstacles, the topological Chern numbers of energy bands have been measured in experiment~\cite{Aidelsburger2015}. 

Having succeded in studying the physics of non-interacting particles in the Harper-Hofstadter model, the natural next step is to explore the realm of strongly interacting systems. The cold atom platforms are well-prepared for this purpose because strong correlations have been induced in many previous cases without synthetic gauge fields. The introduction of strong interaction to photonic crystals and superconducting circuits has also been a long-sought goal, but it has yet to be demonstrated unambiguously in experiments. In the context of Harper-Hofstadter model, Tai {\em et al.} have observed interaction effects in a few body system using quantum gas microscope~\cite{TaiEM2017}. The route to larger systems is still challenging (which is also the case in photonic crystals and superconducting circuits), but we can be cautiously optimistic.

Numerous theoretical studies have been performed to understand strongly interacting particles in optical lattices with synthetic gauge fields. For two-dimensional Harper-Hofstadter models, the low-lying energy bands can be topologically nontrivial and very flat if the flux per plaquette is chosen properly, which enables the realization of fractional quantum Hall states~\cite{WuYH2012-2,Moller2015,Motruk2016,Gerster2017,DongXY2018,Rosson2019}. An opposite limit is the ladder geometry where the system is extended in one direction but only contains two or three legs in the other direction. This quasi-1D setting also harbors a large variety of quantum phases of bosons or fermions~\cite{Grusdt2014,Keles2015,Piraud2015,Cornfeld2015,Petrescu2015,DiDio2015,Barbarino2015,ZengTS1015,Mazza2015,Greschner2015,Greschner2016,SunG2016,Ghosh2017,Calvanese2017,Citro2018}. For one-component bosons, five phases at different filling factors and flux values have been identified: Meissner Mott insulator, Meissner superfluid, vortex Mott insulator, vortex superfluid, and charge density wave~\cite{Piraud2015}. The Meissner and vortex states in the non-interacting limit can be understood in analogy to type II superconductors. For small flux values, there are chiral currents similar to the screening current in the Meissner phase of type II superconductors but no local currents are observed on the rungs. As the flux increases, finite rung currents emerge as in the vortex phase of type II superconductors. 

In this paper, we study quantum phases of bosons with an internal degree of freedom that may assume two possible values (referred to as spin-up and spin-down for simplicity). The magnetic flux per plaquette for the two types of bosons are different in general. If there is no interaction between spin-up and spin-down bosons, we have two independent states for the two components by tuning various parameters. The effect of inter-species interaction on different states is studied numerically using the density matrix renormalization group (DMRG) method. The physics is also analyzed using effective spin theory in certain cases. 

The rest of this paper is organized as follow. In Sec.~\ref{model}, we define the Harper-Hofstadter ladder of two-component bosons and describe our numerical and analytical methods. In Sec.~\ref{result}, we present the results in several cases with different system parameters. The paper is concluded in Sec.~\ref{conclude}.

\section{Models and Methods}
\label{model}

\begin{figure}[ht]
\centering
\includegraphics[width=0.48\textwidth]{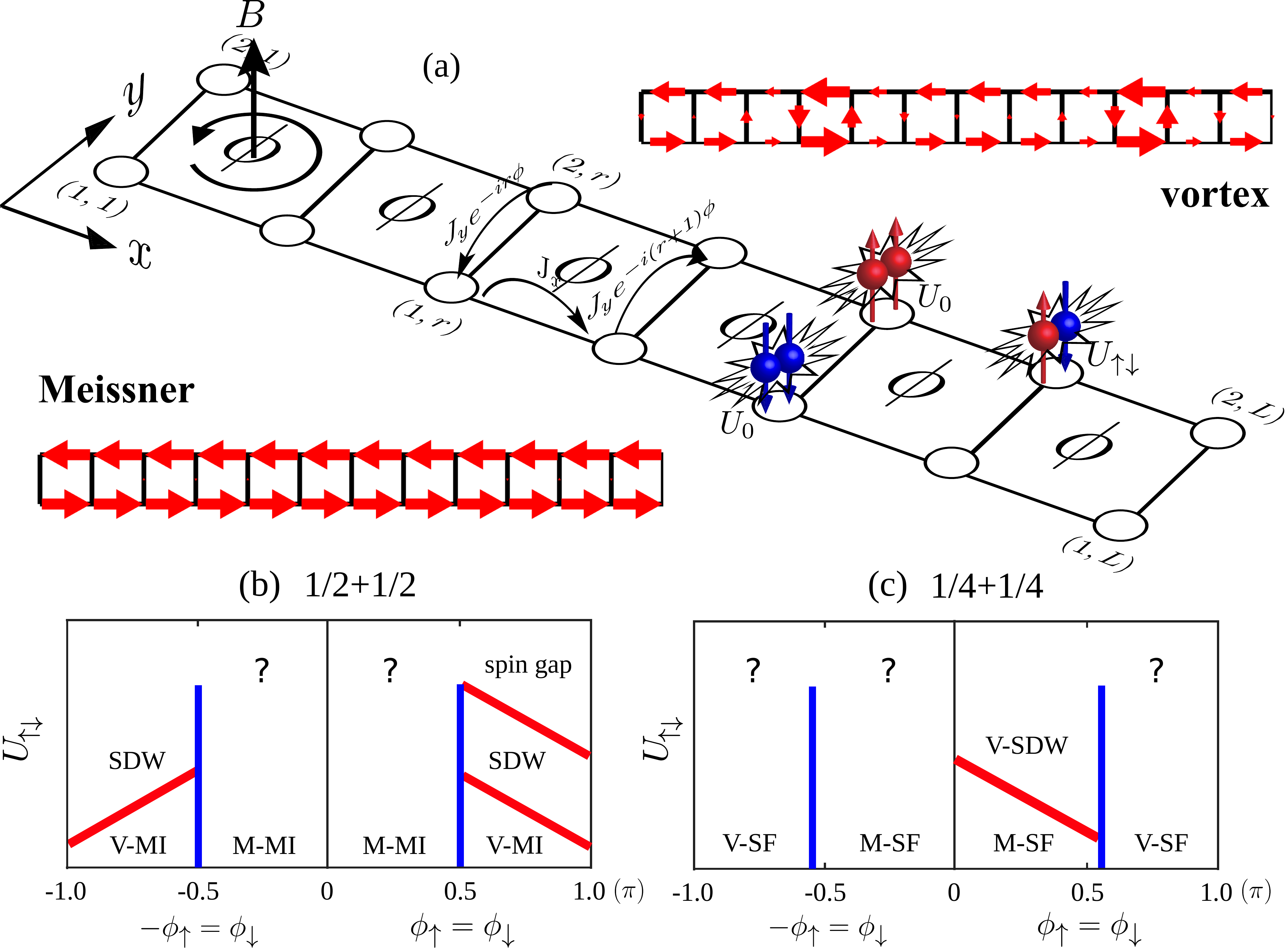}
\caption{(a) Schematics of two-leg Harper-Hofstadter ladder with two-component bosons. The red (blue) dots represent spin-up (spin-down) bosons. The synthetic magnetic fields for the two components can have the same or opposite directions as reflected by the hopping phases along the $y$ direction. The absolute value of the flux in each plaquette is a rational number $\phi=2{\pi}p/q$. $U_{0}$ and $U_{\uparrow\downarrow}$ denote intra- and inter-species interaction strengths. Two representative current patterns in the Meissner and vortex phases are also shown. (b) Sketchy phase diagram at filling factors $\nu_{\uparrow}=\nu_{\downarrow}=1/2$. (c) Sketchy phase diagram at filling factors $\nu_{\uparrow}=\nu_{\downarrow}=1/4$. The acronyms are Meissner Mott insulator (M-MI), votex Mott insulator (V-MI), Meissner superfluid (M-SF), vortex superfluid (V-SF), spin density wave (SDW), vortex spin density wave (V-SDW). The question marks indicate that no phase transition is observed even for very large $U_{\uparrow\downarrow}$.}
\label{fig:scheme}
\end{figure}

The system that we study in this paper is depicted schematically in Fig.~\ref{fig:scheme}. There are two legs labeled as $l=1,2$ and $L$ rungs indexed by $r\in[1,2,\ldots,L]$. The system contains two types of bosons called spin-up and spin-down, which experience either the same or opposite synthetic magnetic fluxes. The absolute value of the flux in each plaquette is a rational number $\phi=2{\pi}p/q$. The creation (annihilation) operators for the bosons are $b^{\dag}_{\sigma,l,r}$ with $\sigma=\uparrow,\downarrow$. The Harper-Hofstadter model for two-component bosons is
\begin{eqnarray}
H &=& - t_{x} \sum_{\sigma=\uparrow,\downarrow} \sum_{l=1,2} \sum_{r} \left( a^{\dag}_{\sigma,l,r} a_{\sigma,l,r+1} + {\rm H.c.} \right) \nonumber \\ 
  &\phantom{=}& - t_{y} \sum_{\sigma=\uparrow,\downarrow} \sum_{r} \left( e^{-ir\phi_{\sigma}} a^{\dag}_{\sigma,1,r} a_{\sigma,2,r} + {\rm H.c.} \right) \nonumber \\ 
  &\phantom{=}& + \frac{U_{0}}{2} \sum_{\sigma=\uparrow,\downarrow} \sum_{l=1,2} \sum_{r} n_{\sigma,l,r} (n_{\sigma,l,r}-1) \nonumber \\ 
  &\phantom{=}& + U_{\uparrow\downarrow} \sum_{l=1,2} \sum_{r} n_{\uparrow,l,r} n_{\downarrow,l,r},
\label{eq:model}
\end{eqnarray}
where $n_{\sigma,l,r}=a^{\dag}_{\sigma,l,r} a_{\sigma,l,r}$ is the particle number operator, $U_{0}$ is the intra-species onsite repulsion, and $U_{\uparrow\downarrow}$ is the inter-species onsite repulsion. The total number of bosons in each component are denoted as $N_{\sigma}$ and the filling factor is defined as $\nu_{\sigma}=N_{\sigma}/2L$. If the two components have the same magnetic field, $\phi_{\uparrow}=\phi_{\downarrow}=\phi$. If the two components have opposite magnetic field, $\phi_{\uparrow}=\phi$ and $\phi_{\downarrow}=-\phi$. The model has an SU(2) symmetry when $\phi_{\uparrow}=\phi_{\downarrow}$ and $U_{0}=U_{\uparrow\downarrow}$. In other cases, the model has a ${\mathbb Z}_{2}$ symmetry that corresponds to interchange of the two types of bosons (and change the signs of the hopping phases if they are opposite). 

The system with various choices of parameters are studied using the density matrix renormalization group (DMRG) method~\cite{White1992,Schollwock2011,Hubig2015}. This algorithm variationally searches for the ground state within the class of matrix product states (MPS). It is best suited for open boundary conditions as we will adopt throughout this paper. If the basis states for individual lattice sites are denoted $\{|s_{i}\rangle\}$, a generic MPS has the form
\begin{eqnarray}
|\psi\rangle = \sum_{s_{1}} \ldots \sum_{s_{L}} B^{s_{1}}_{1} B^{s_{2}}_{2} \ldots B^{s_{L}}_{L}  |s_{1},s_{2},\ldots,s_{L}\rangle,
\end{eqnarray}
where the $B^{s_{i}}_{i}$'s are matrices to be optimized iteratively by sparse matrix eigensolver. The bond dimension $D$ is defined as the maximal dimension of the $B^{s_{i}}_{i}$ matrices. The computational resource needed for DMRG calculations is related to the bipartite von Neumann entanglement entropy (EE). For one-dimensional gapped system with short-range interactions, the EE is bounded and the $D$ required for accurate simulation does not need to increase with the system size~\cite{Verstraete2006,Hastings2007}. In contrast, $D$ needs to grow with the system size if the system is gapless because EE exhibits a logarithmic growth~\cite{Calabrese2004}. The maximal bond dimension that we have used in this paper is $D=6000$. This produces accurate results as quantified by the energy variance $\langle\psi|H^{2}|\psi\rangle - (\langle\psi|H|\psi\rangle)^{2}$ that falls in the range of $10^{-4}{\sim}10^{-7}$. For certain choices of parameters, the system can be studied using standard analytical perturbative methods~\cite{Auerbach1994}. This helps us to understand the superexchange process between lattice sites and produces effective spin models where the spins correspond to singly-occupied lattice sites.

\section{Numerical Results}
\label{result}

The Hamiltonian contains five parameters $t_{x},t_{y},\phi_{\sigma},U_{0},U_{\uparrow\downarrow}$. For each set of parameters, one may study various filling factors and choose the magnetic fluxes to be the same or different for the two components. This makes it rather difficult to perform an exhaustive study of the system, so we shall focus on several specific choices of parameters that are motivated by known results in one-component systems~\cite{Greschner2015,Greschner2016}. The hopping parameters $t_{x}$ and $t_{y}$ are both fixed at $1$. The intra-species interaction strength $U_{0}$ is chosen to be infinite, which forbids the presence of more than one boson with the same spin on any lattice site. All numerical results quoted below are for the $L=60$ system. 

\subsection{vortex Mott insulator and the same magnetic field}

One representative phase of one-component hard-core bosons is the gapless vortex Mott insulator at filling factor $1/2$ and $\pi/2{\lesssim}\phi{\lesssim}3\pi/2$~\cite{Greschner2015,Greschner2016}. To understand this name properly, we need to distinguish between two different gaps. The mass gap is defined as
\begin{eqnarray}
{\Delta E}_{\rm ma} = \frac{1}{2} \left[ E_{\rm gs}(N+1) + E_{\rm gs}(N-1) \right] - E_{\rm gs}(N)
\end{eqnarray}
and the excitation gap is defined as
\begin{eqnarray}
{\Delta E}_{\rm ex} = E_{\rm ex}(N) - E_{\rm gs}(N),
\end{eqnarray}
where $E_{\rm gs}(N)$ [$E_{\rm ex}(N)$] is the ground state (first excited state) energy in the subspace with $N$ bosons. The state qualifies as a Mott insulator since there is a mass gap but it is called gapless as the excitation gap vanishes. This phase features an average rung current and the scaling of its von Neumann EE gives central charge $1$. For concreteness, we focus on the case with $\nu_{\uparrow}=\nu_{\downarrow}=1/2$ and $\phi_{\uparrow}=\phi_{\downarrow}=4\pi/5$. If the inter-species interaction $U_{\uparrow\downarrow}$ is zero, the system would simply be two independent gapless vortex Mott insulators. Numerical results suggest that two phase transitions occur as $U_{\uparrow\downarrow}$ increases.

\begin{figure}[ht]
\centering
\includegraphics[width=0.48\textwidth]{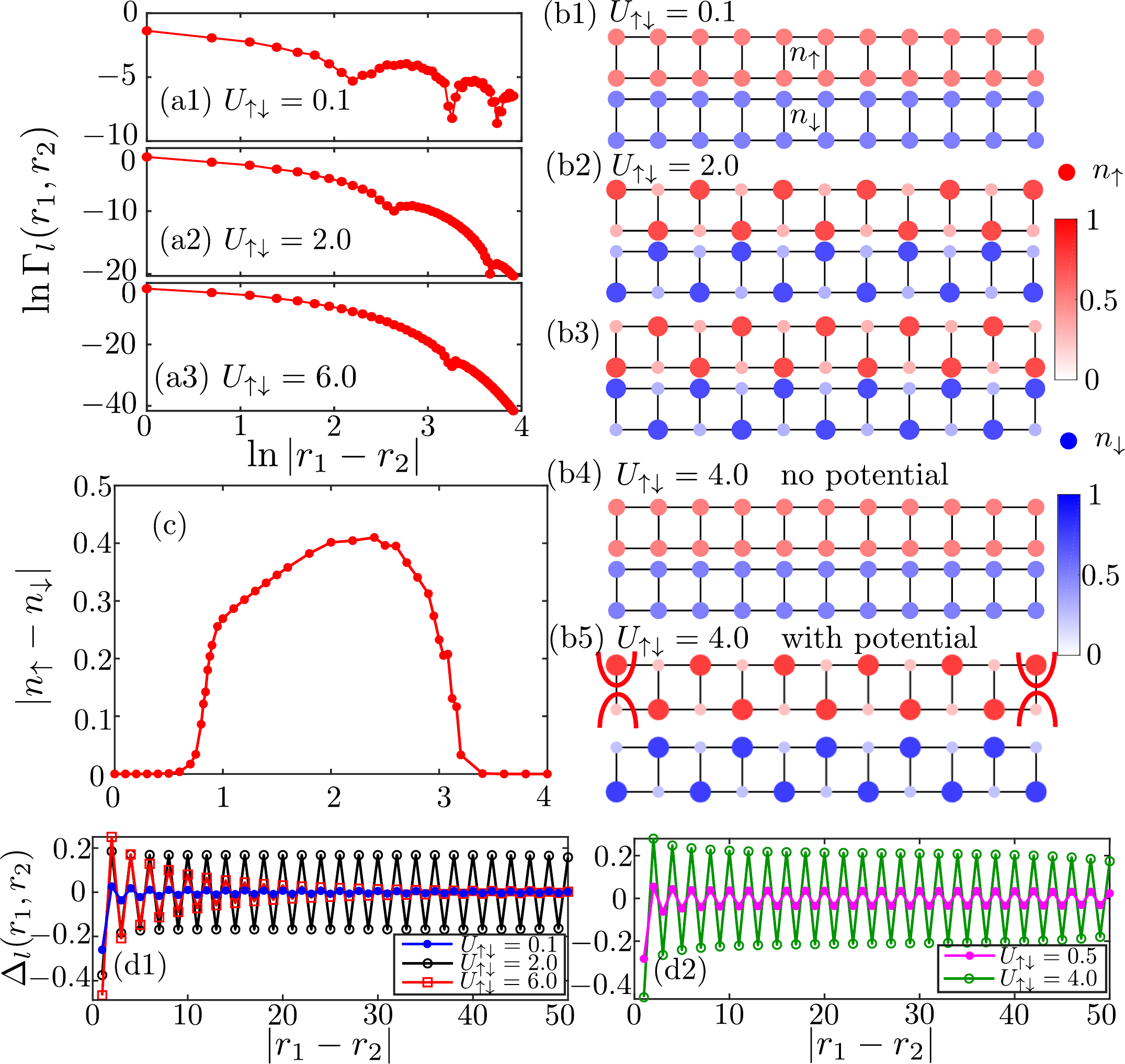}
\caption{Numerical results at filling factor $1/2+1/2$ where the two types of bosons have the same magnetic field. (a) The charge correlator $\Gamma_{l}(r_{1},r_{2})$ on the first leg. (b1,b2,b4,b5) The particle density profiles for three choices of $U_{\uparrow\downarrow}$. The red curves in (b5) represent the boundary pinning potential $(-1)^{l+r} n_{\uparrow,l,r}$ on the first and last rungs. (b3) The particle density profile of another state that is degenerate with the one in (b2). In each subpanel, the top (bottom) one is for spin-up (spin-down) bosons. (c) The absolute difference of particle numbers on the first leg. (d1-d2) The density difference correlator $\Delta_{l}(r_{1},r_{2})$ on the first leg.}
\label{fig:12+12A1}
\end{figure}

To begin with, we probe the system using the charge correlator 
\begin{eqnarray}
\Gamma_{\sigma,l}(r_{1},r_{2})=\langle a^{\dag}_{\sigma,l,r_{1}} a_{\sigma,l,r_{2}}\rangle,
\label{eq:ChargeCorrelator}
\end{eqnarray}
which helps us to distinguish between superfluidity and other phases. The Mermin-Wagner theorem dictates that there is no true long-range order in one-dimensional systems with short-range interactions~\cite{Mermin1966,Hohenberg1967}. Instead, the superfluid state has a quasi-long-range order and the correlator decays algebraically as
\begin{eqnarray}
\Gamma_{\sigma,l}(r_{1},r_{2}) = f\cos(q|r_{1}-r_{2}|) |r_{1}-r_{2}|^{-\alpha}.
\label{eq:SuperFluidOrder}
\end{eqnarray}
In contrast, the Mott insulating state has no quasi-long-range superfluid order, so the correlator decays exponentially with distance as
\begin{eqnarray}
\Gamma_{\sigma,l}(r_{1},r_{2}) = f\cos(q|r_{1}-r_{2}|) e^{-\alpha|r_{1}-r_{2}|}.
\label{eq:MottInsulatorOrder}
\end{eqnarray}
For the whole range of $U_{\uparrow\downarrow}$ that we have checked, the charge correlator $\Gamma_{\sigma,l}(r_{1},r_{2})$ for spin-up bosons on the first leg decays quickly so there is not quasi-long-range superfluid order [see Fig.~\ref{fig:12+12A1} (a) for the cases with $U_{\uparrow\downarrow}=0.1,2.0,6.0$]. 

The expectation values of the particle number operator $n_{\sigma,l,r}=a^{\dag}_{\sigma,l,r}a_{\sigma,l,r}$ have also been computed. For small $U_{\uparrow\downarrow}$, the two types of bosons spread evenly on the lattice sites as shown in Fig.~\ref{fig:12+12A1} (b1). As $U_{\uparrow\downarrow}$ passes $\sim 0.9$, a clear density modulation emerges in Fig.~\ref{fig:12+12A1} (b2): if one lattice site has $n_{\uparrow}<0.5<n_{\downarrow}$, then its neighbor has $n_{\downarrow}<0.5<n_{\uparrow}$. If $U_{\uparrow\downarrow}$ further increases to $\sim 3.4$, the particle density profile becomes uniform again. The absolute difference of particle numbers on the first leg, which is defined as
\begin{eqnarray}
\left| n_{\uparrow}-n_{\downarrow} \right| = \frac{1}{L} \sum_{r} \left| n_{1,r,\uparrow} - n_{1,r,\downarrow} \right|,
\end{eqnarray}
displays two apparent changes as shown in Fig.~\ref{fig:12+12A1} (c). Naively, these observations appear to indicate that there are two phase transitions at $U_{\uparrow\downarrow} \sim 0.9$ and $3.4$, respectively. However, a closer inspection demonstrates that the phase transitions actually occur at $U_{\uparrow\downarrow} \sim 0.3$ and $\sim 5.0$.

The ${\mathbb Z}_{2}$ symmetry that corresponds to interchange of the two types of bosons is broken in the density modulated phase. In fact, if there is a state with the density profile of Fig.~\ref{fig:12+12A1} (b2) ($n_{\downarrow,1,1}<0.5<n_{\uparrow,1,1},n_{\uparrow,2,1}<0.5<n_{\downarrow,2,1}$ etc.), it is degenerate with another state that has exactly opposite density profile shown in Fig.~\ref{fig:12+12A1} (b3) ($n_{\uparrow,1,1}<0.5<n_{\downarrow,1,1},n_{\downarrow,2,1}<0.5<n_{\uparrow,2,1}$ etc.). The two degenerate states may be resolved by DMRG in some cases, but it is also possible that numerics fail to distinguish them, in which case the result is a superposition of them and there is no explicit density modulation. The presence of ${\mathbb Z}_{2}$ symmetry breaking can be probed more accurately using the correlator
\begin{eqnarray}
\Delta_{l}(r_{1},r_{2}) = \langle \delta n_{l,r_{1}} \delta n_{l,r_{2}} \rangle
\end{eqnarray}
of the density difference $\delta n_{l,r} = n_{\uparrow,l,r}-n_{\downarrow,l,r}$. The results presented below are for the first leg, but using the other leg would lead to the same conclusion. Because the ${\mathbb Z}_{2}$ symmetry is a discrete one, the Mermin-Wagner theorem does not preclude the existence of true long-range order. As shown in Fig.~\ref{fig:12+12A1} (d1) and (d2), the long-range correlation of the density difference is evident at $U_{\uparrow,\downarrow}=0.5,2.0,4.0$ but is absent at $U_{\uparrow,\downarrow}=0.1,6.0$. By comparing $\Delta_{l}(r_{1},r_{2})$ and the density profile, we conclude that the DMRG calculations did not differentiate the two degenerate states when $0.3 \lesssim U_{\uparrow\downarrow} \lesssim 0.9$ and $3.4 \lesssim U_{\uparrow\downarrow} \lesssim 5.0$. To further corroborate this interpretation~\cite{Stoudenmire2012}, we add a boundary pinning potential $(-1)^{l+r} n_{\uparrow,l,r}$ on the first and last rungs such that spin-up bosons are attracted or repelled and the ${\mathbb Z}_{2}$ symmetry is explicitly broken~\footnote{Except for those explicitly stated, the numerical results presented in this paper are for the cases without the boundary pinning potential.}. This term results in a density modulation at $U_{\uparrow\downarrow}=4$ but does not cause such changes at $U_{\uparrow\downarrow} \lesssim 0.3$ or $\gtrsim 5.0$. The effect of boundary pinning potential can also be seen in the entanglement entropy (see below).

\begin{figure}[ht]
\centering
\includegraphics[width=0.48\textwidth]{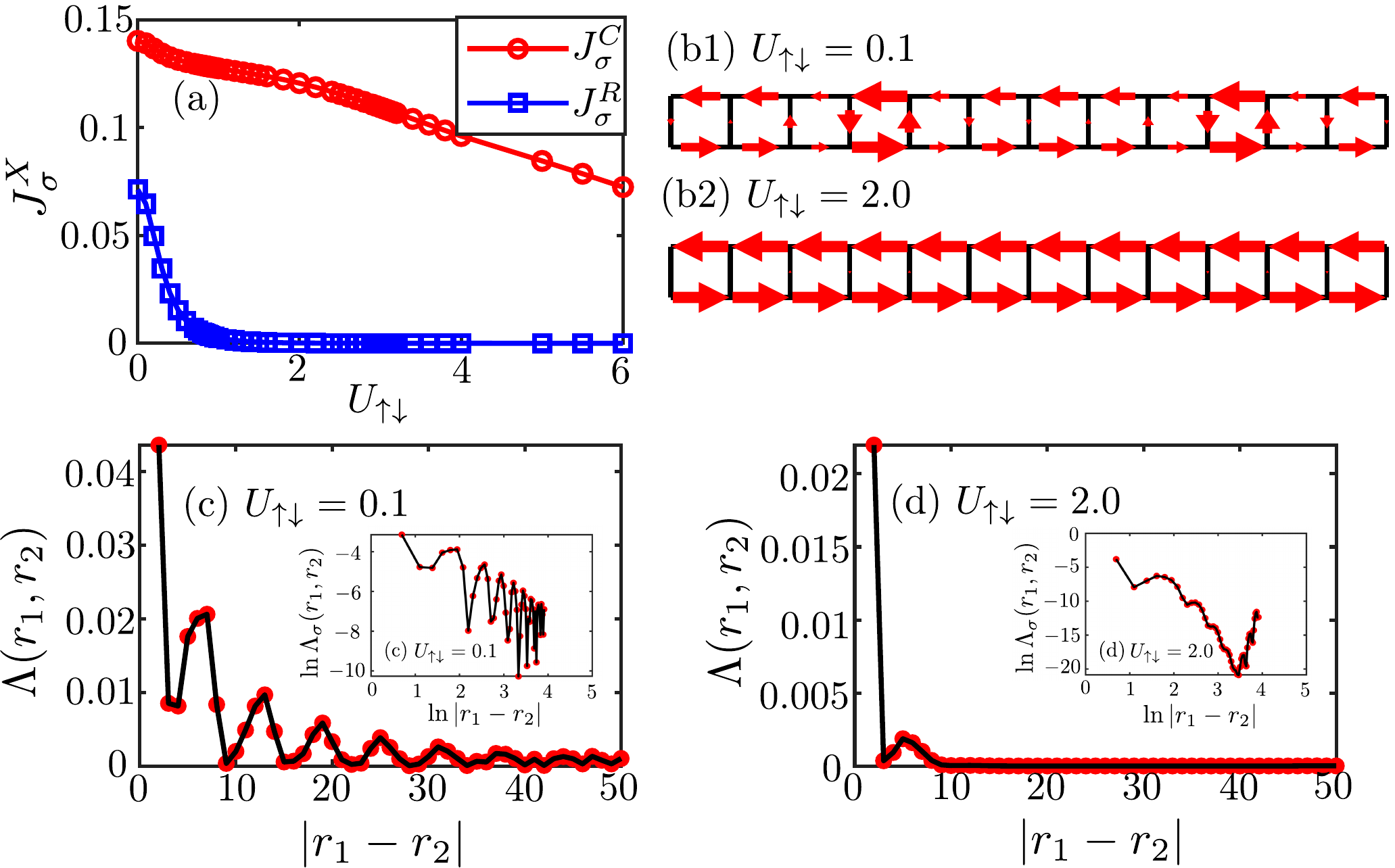}
\caption{Numerical results at filling factor $1/2+1/2$ where the two types of bosons have the same magnetic field. (a) The chiral current and the average rung current of the spin-up bosons. (b1-b2) The current pattern of the spin-up bosons. (c-d) The rung current correlator $\Lambda_{\sigma}(r_{1},r_{2})$ of the spin-up bosons. The insets show the results in log-log scale.}
\label{fig:12+12A2}
\end{figure}

The local current patterns have different features in the Meissner and vortex phases. In the Heisenberg equation of motion for the particle number, we take expectation value on both sides to yield
\begin{eqnarray}
\left\langle \frac{dn_{\sigma,l,r}}{dt} \right\rangle = i \langle \left[ H,n_{\sigma,l,r} \right] \rangle.
\end{eqnarray}
On the other hand, the changing rate of $n_{\sigma,l,r}$ is determined by the current flowing out from site
\begin{eqnarray}
\left\langle \frac{dn_{\sigma,l,r}}{dt} \right\rangle = - \sum_{(l',r')} j_{\sigma}[(l',r')\rightarrow(l,r)],
\end{eqnarray}
This allows us to compute current operators using the commutator $[H,n_{\sigma,l,r}]$. Based on this analysis, one can define a leg current
\begin{eqnarray}
j^{\parallel}_{\sigma,l,r} = it_{x} \left( a^{\dagger}_{\sigma,l,r+1} a_{\sigma,l,r} - a^{\dagger}_{\sigma,l,r} a_{\sigma,l,r+1} \right)
\end{eqnarray}
and a rung current
\begin{eqnarray}
j^{\perp}_{\sigma,r} = i t_{y} \left( e^{-ir\phi_{\sigma}} a^{\dagger}_{\sigma,1,r} a_{\sigma,2,r} - e^{ir\phi_{\sigma}} a^{\dagger}_{\sigma,2,r} a_{\sigma,1,r} \right).
\end{eqnarray}
It is useful to define a chiral current 
\begin{eqnarray}
j^{C}_{\sigma} = \frac{1}{2L} \sum_{r} \left\langle j^{\parallel}_{\sigma,1,r}- j^{\parallel}_{\sigma,2,r} \right\rangle
\end{eqnarray}
to characterize the current encircling the ladder and an average rung current 
\begin{eqnarray}
j^{R}_{\sigma} = \frac{1}{L} \sum_{r} \left| \langle j^{\perp}_{\sigma,r} \rangle \right|
\end{eqnarray}
to characterize interchain current flow. In the vortex phase, the average rung current is nonzero and there are many possible vortex configurations. In the Meissner phase, the average rung current vanishes but a finite chiral current is present.

The chiral current $J^{C}_{\sigma}$ and average rung current $J^{R}_{\sigma}$ are presented in Fig.~\ref{fig:12+12A2}. The spin-up component is used here, but the spin-down component has no difference due to the ${\mathbb Z}_{2}$ symmetry. As the system passes the first phase transition at $U_{\uparrow\downarrow}{\sim}0.9$, $J^{R}_{\sigma}$ decreases to zero but the chiral current remains finite, yet no qualitative changes were observed at the purported second phase transition [Fig.~\ref{fig:12+12A2} (a)]. Two current patterns at $U_{\uparrow\downarrow}=0.1$ and $U_{\uparrow\downarrow}=2.0$ are shown in Fig.~\ref{fig:12+12A2} (b), where one can clearly see vortex structures and chiral currents, respectively. Furthermore, the rung current correlator 
\begin{eqnarray}
\Lambda_{\sigma}(r_{1},r_{2}) = \langle j^{\perp}_{\sigma,r_{1}} j^{\perp}_{\sigma,r_{2}} \rangle
\end{eqnarray}
exhibits an algebraic decay at $U_{\uparrow\downarrow}=0.1$ [Fig.~\ref{fig:12+12A2} (c)] but an exponential decay at $U_{\uparrow\downarrow}=2.0$ [Fig.~\ref{fig:12+12A2} (d)]. These behaviors indicate that the system transits from a vortex phase to a Meissner phase as $U_{\uparrow\downarrow}$ increases. The two phases with different density profiles at $U_{\uparrow\downarrow}{\gtrsim}0.3$ are not different in their current patterns.

\begin{figure}[ht]
\centering
\includegraphics[width=0.48\textwidth]{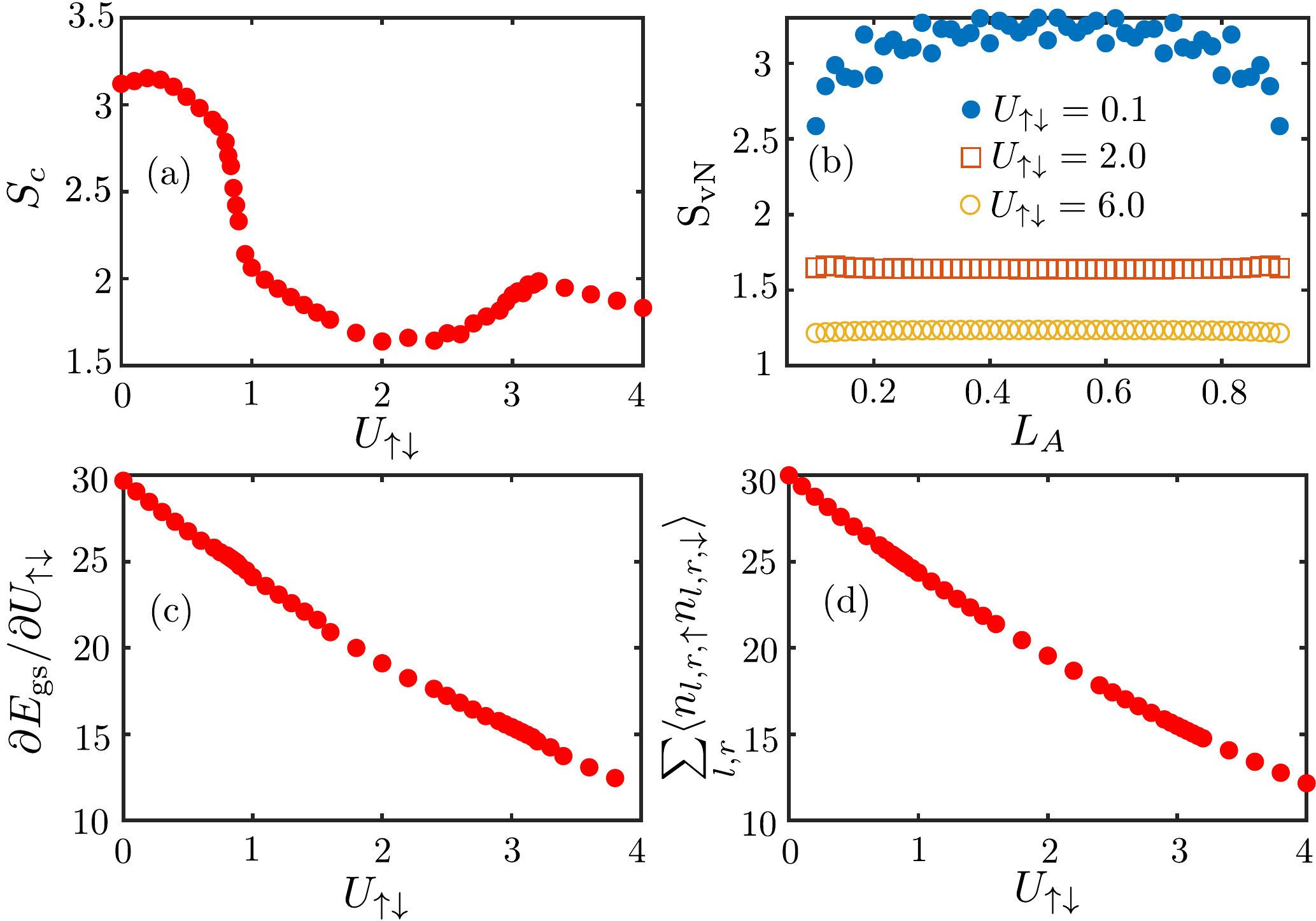}
\caption{Numerical results at filling factor $1/2+1/2$ where the two types of bosons have the same magnetic field. (a) The von Neumann EE at the center of the system. (b) The von Neumann EE versus subsystem size. The fitting parameters at $U_{\uparrow\downarrow}=0.1$ using Eq.~(\ref{eq:Cardy}) are $c=2.747$ and $g=1.930$. (c) The derivative of the ground state energy with respect to $U_{\uparrow\downarrow}$. (d) The expectation value $\sum\limits_{l,r} \langle n_{\uparrow,l,r} n_{\downarrow,l,r} \rangle$.}
\label{fig:12+12A3}
\end{figure}

\begin{table}[ht]
\begin{tabular}{cccc}
\hline
\hline
$L_{A}$ & 30 & 40 \\
\hline
$U_{\uparrow\downarrow}=0.5$ (N) & 3.0452 & 3.0384 \\
$U_{\uparrow\downarrow}=0.5$ (P) & 2.2323 & 2.2329 \\
$U_{\uparrow\downarrow}=4.0$ (N) & 1.8324 & 1.8157 \\
$U_{\uparrow\downarrow}=4.0$ (P) & 1.1723 & 1.1718 \\
$U_{\uparrow\downarrow}=6.0$ (N) & 1.2367 & 1.2360 \\
$U_{\uparrow\downarrow}=6.0$ (P) & 1.2362 & 1.2354 \\
\hline
\hline
\end{tabular}
\caption{The entanglement entropy at filling factor $1/2+1/2$ where the two types of bosons have the same magnetic field. The P (N) in parenthesis means that boundary pinning potential is (not) added. The presence of this potential substantially reduces the entanglement entropy in the $\mathbb{Z}_{2}$ symmetry breaking phase ($U_{\uparrow\downarrow}=0.5,4.0$) but not in the large $U_{\uparrow\downarrow}$ phase.}
\label{Table1}
\end{table}

The bipartition von Neumann EE is very useful for probing quantum phases. The system is divided into two subsystems $A$ and $B$ which have the first $L_{A}$ rungs and the other rungs, respectively. The reduced density matrix $\rho_{A}$ of the $A$ subsystem is computed by tracing out part $B$ and the von Neumann EE is defined as 
\begin{eqnarray}
S_{\rm vN} =-{\rm Tr}(\rho_{A}\ln\rho_{A}).
\label{eq:vNEE}
\end{eqnarray}
The functional form of the von Neumann EE can be used to check whether the system is gapless or gapped. If the low-energy physics of a system (with open boundary condition) is described by a conformal field theory (CFT), the von Neumann EE has the scaling form
\begin{eqnarray}
S_{\rm vN} = \frac{c}{6} \ln\left[ \frac{L}{\pi} \sin\frac{L_A}{L} \pi \right] + g + \cdots,
\label{eq:Cardy}
\end{eqnarray}
where $c$ is the central charge of the CFT, $g$ is a constant, and $\cdots$ represent other non-universal terms~\cite{Calabrese2004}. In contrast, the von Neumann EE saturates to a constant value in the bulk of a system when it is gapped (there is no CFT description for such cases). The von Neumann EE at the center of the ladder exhibits two discontinuities as shown in Fig.~\ref{fig:12+12A3} (a). This is consistent with our previous conclusion that there should be two phase transitions. For the two phases at intermediate and large $U_{\uparrow\downarrow}$, the von Neumann EE quickly saturates to constant values as the subsystem size $L_{A}$ increases [Fig.~\ref{fig:12+12A3} (b)], so they should be gapped phases. The strong oscillation in the EE at very small $U_{\uparrow\downarrow}$ makes it difficult to extract the central charge, but direct fitting at $U_{\uparrow\downarrow}=0.1$ gives a value $2.7$ that is basically consistent with the expected value $2$. In previous works~\cite{Greschner2015,Greschner2016}, it has been shown that the low-energy theory for the one-component system at half filling is a CFT with $c=1$. This means that we have a $c=2$ CFT in the the two-component system if $U_{\uparrow\downarrow}=0$. If a small but finite $U_{\uparrow\downarrow}$ does not gap out any degrees of freedom, the system should still be described by the same $c=2$ CFT. The entanglement entropy also provides further support for our previous analysis about the $\mathbb{Z}_2$ symmetry breaking phase. If the boundary pinning potential is present, the compute program is able to pick out one state from two degenerate states. In contrast, a superposition of the two degenerate states are obtained if there is no boundary pinning potential. It is expected that the entanglement entropy in the former cases is considerably lower than that in the latter cases, which is confirmed by the numerical results in Table~\ref{Table1}.

The derivative of the ground state energy with respect to $U_{\uparrow\downarrow}$ is shown in Fig.~\ref{fig:12+12A3} (c). It can actually be computed using the Hellmann-Feynman theorem as
\begin{eqnarray}
\dfrac{\partial E_{\rm gs}}{\partial U_{\uparrow\downarrow}} = \sum\limits_{l,r} n_{\uparrow,l,r} n_{\downarrow,l,r}
\end{eqnarray}
and the result is also shown in Fig.~\ref{fig:12+12A3} (c). The absence of singularities implies that the two phase transitions are continous.

The density modulated state observed at intermediate $U_{\uparrow\downarrow}$ should be a spin density wave (SDW). An intuitive picture for this state is that $U_{\uparrow\downarrow}$ disfavors the configurations with two bosons occupying the same lattice sites. The superexchange process between lattice sites select an antiferromagnetic order in which two neighboring sites are occupied by opposite spins. This gives rise to the density modulation as manifested by oscillations of ${\delta}n_{l,r}$. However, it is interesting that this picture does not persist as $U_{\uparrow\downarrow}\rightarrow+\infty$. In the large $U_{\uparrow\downarrow}$ limit, the low-energy subspace of the model has exactly one boson per site. In this subspace, the effective degrees of freedom are spin operators denoted as ${\mathbf S}_{l,r}$, which are described by the Hamiltonian
\begin{eqnarray}
H_{\rm spin} = J \sum_{l,r} {\mathbf S}_{l,r} \cdot {\mathbf S}_{l,r+1} + J \sum_{r} {\mathbf S}_{1,r}\cdot {\mathbf S}_{2,r+1}.
\label{eq:spin}
\end{eqnarray}
with $J=4/U_{\uparrow\downarrow}$. It is well-known that this spin ladder has a spin gap~\cite{Shelton1996}, which is consistent with our results at large $U_{\uparrow\downarrow}$. The Appendix provides more details about how to derive this Hamiltonian.

\subsection{vortex Mott insulator and opposite magnetic fields}

\begin{figure}[ht]
\centering
\includegraphics[width=0.48\textwidth]{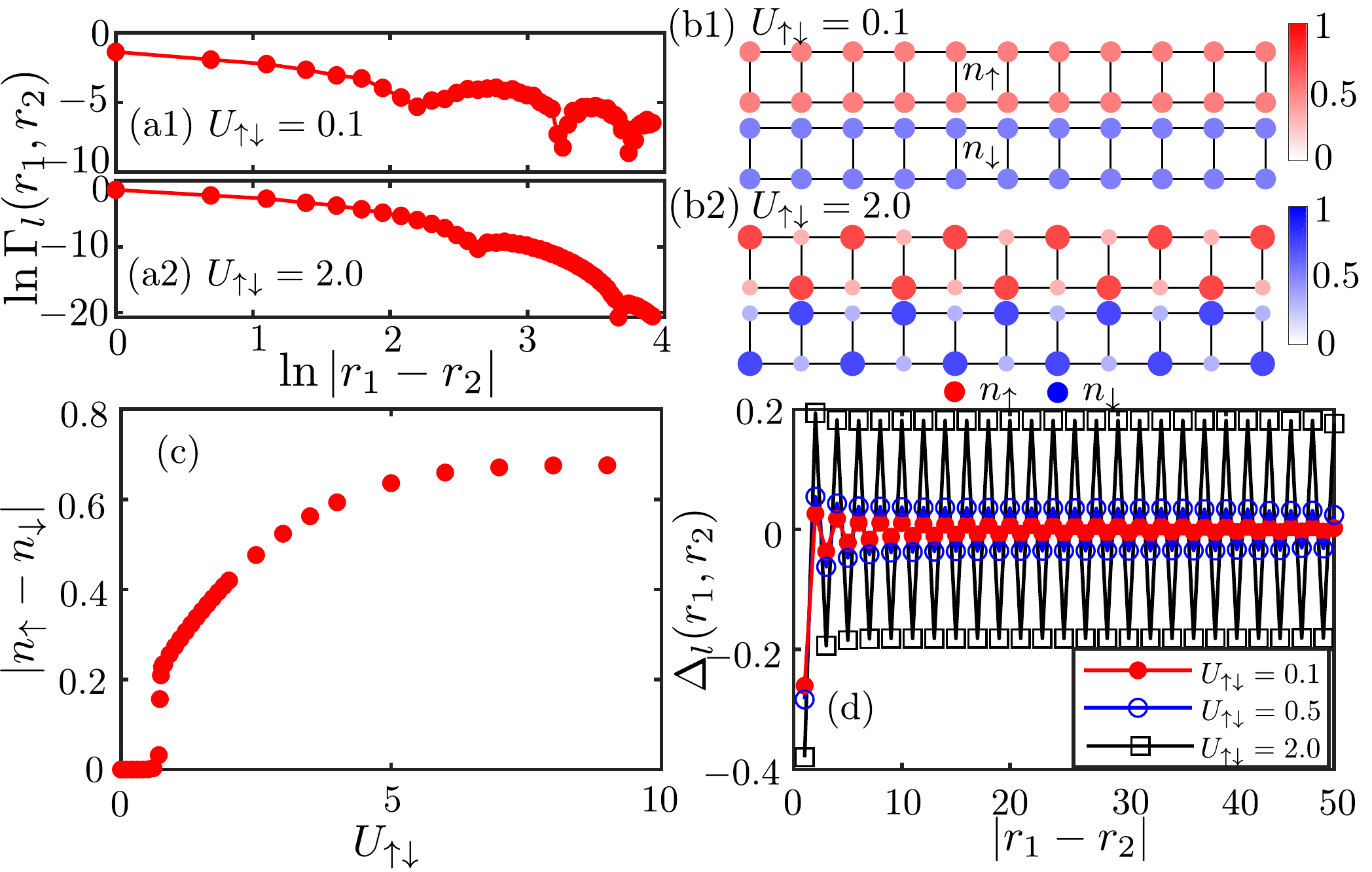}
\caption{Numerical results at filling factor $1/2+1/2$ where the two types of bosons have opposite magnetic fields. (a) The charge correlator $\Gamma_{l}(r_{1},r_{2})$ on the first leg. (b1-b2) The particle density profiles. In each subpanel, the top (bottom) one is for spin-up (spin-down) bosons. (c) The absolute difference of particle numbers on the first leg. (d) The density difference correlator $\Delta_{l}(r_{1},r_{2})$ on the first leg.}
\label{fig:12+12B1}
\end{figure}

\begin{figure}[ht]
\centering
\includegraphics[width=0.48\textwidth]{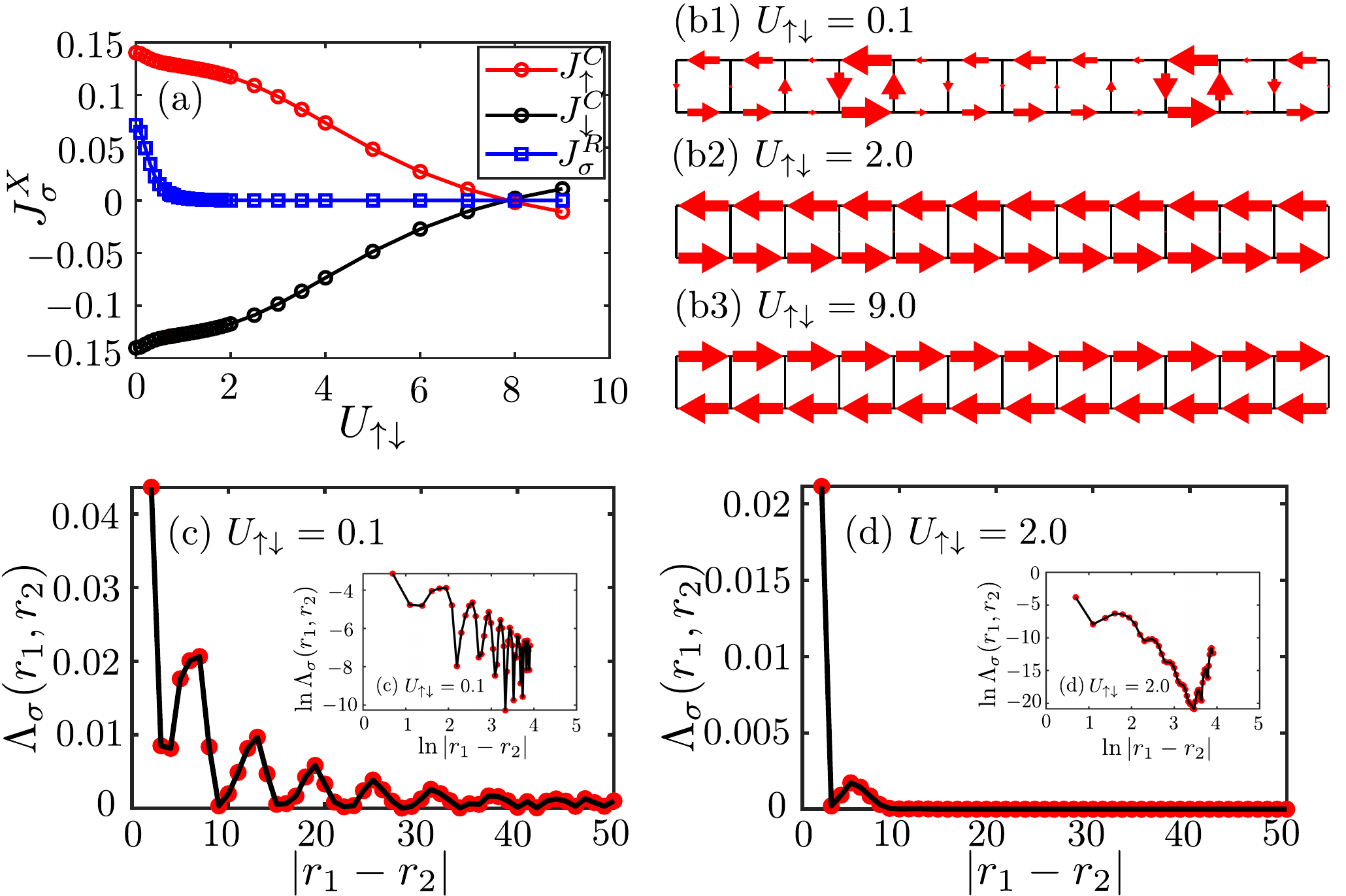}
\caption{Numerical results at filling factor $1/2+1/2$ where the two types of bosons have opposite magnetic fields. (a) The chiral current of both types of bosons and the average rung current of the spin-up bosons. (b1-b3) The current pattern of the spin-up bosons. (c-d) The rung current correlator $\Lambda_{\sigma}(r_{1},r_{2})$ of the spin-up bosons. The insets show the results in log-log scale.}
\label{fig:12+12B2}
\end{figure}

\begin{figure}[ht]
\centering
\includegraphics[width=0.48\textwidth]{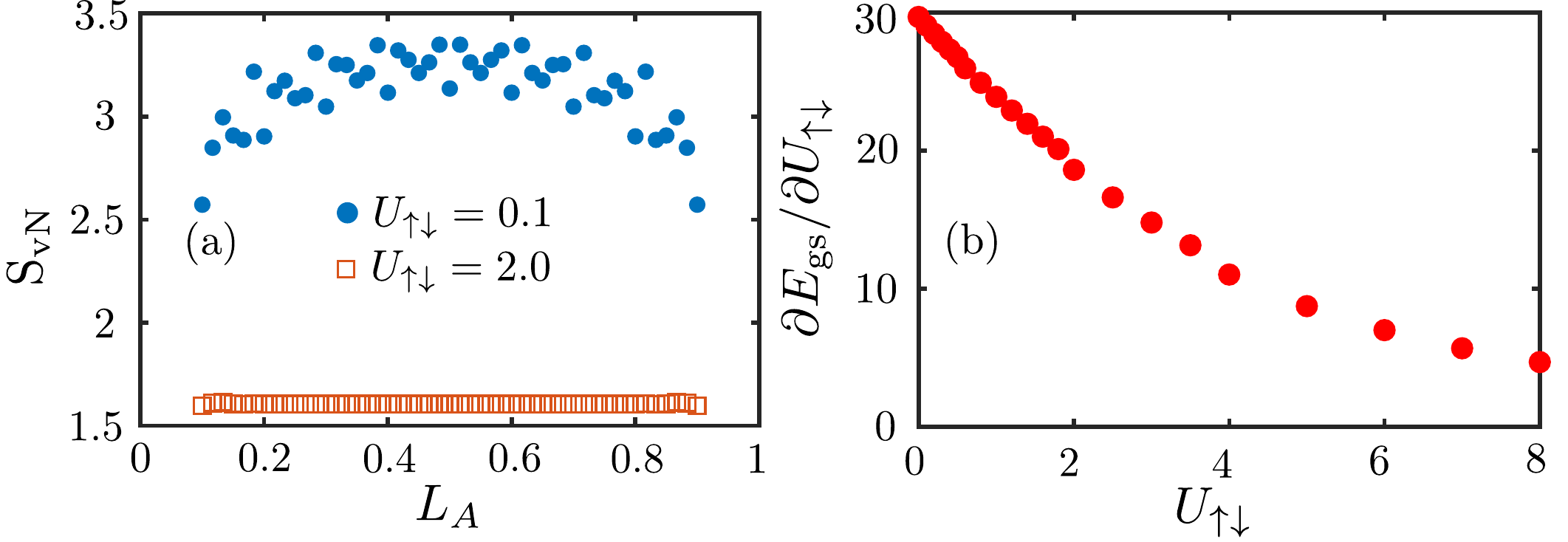}
\caption{Numerical results at filling factor $1/2+1/2$ where the two types of bosons have opposite magnetic fields. (a) The von Neumann EE versus subsystem size. The fitting parameters at $U_{\uparrow\downarrow}=0.1$ using Eq.~(\ref{eq:Cardy}) are $c=2.615$ and $g=1.979$. (b) The derivative of the ground state energy with respect to $U_{\uparrow\downarrow}$.}
\label{fig:12+12B3}
\end{figure}

The difference between this case and the previous one is that the magnetic field for the spin-down bosons is reversed. The system still has a ${\mathbb Z}_{2}$ symmetry, but it corresponds to exchange of the two types of bosons and reversal of their respective magnetic fields at the same time. Numerical results suggest that there is one phase transition as $U_{\uparrow\downarrow}$ increases. The state at large $U_{\uparrow\downarrow}$ is a gapped Meissner phase that breaks the ${\mathbb Z}_{2}$ symmetry.

The charge correlator on the first leg as shown in Fig.~\ref{fig:12+12B1} (a) also indicates the absence of any superfluid order. The particle density profile is uniform at small $U_{\uparrow\downarrow}$ but it gets modulated when $U_{\uparrow\downarrow} \gtrsim 0.9$ [Fig.~\ref{fig:12+12B2} (b)] and remains so up to the largest value that we have checked. This is also reflected in the absolute difference of particle numbers on the first leg, which changes only once from zero to nonzero as shown in Fig.~\ref{fig:12+12B1} (c). As in the previous subsection, the density difference correlator $\Delta_{l}(r_{1},r_{2})$ on the first leg reveals that the ${\mathbb Z}_{2}$ symmetry is already broken when $U_{\uparrow\downarrow} \gtrsim 0.3$ [Fig.~\ref{fig:12+12B1} (d)]. Based on similar arguments as used before, we believe that the system enters a spin density wave phase at $U_{\uparrow\downarrow} \gtrsim 0.3$. It is interesting to note that DMRG is always able to distinguish the two degenerate states with opposite particle density profiles for large $U_{\uparrow\downarrow}$.

The chiral current $J^{C}_{\sigma}$ and average rung current $J^{R}_{\sigma}$ are presented in Fig.~\ref{fig:12+12B2}. The spin-up and spin-down components have exactly opposite chiral currents as required by the ${\mathbb Z}_{2}$ symmetry. The average rung current vanishes after passing the phase transition. As shown in Fig.~\ref{fig:12+12B2} (b), vortex structures and chiral currents are observed at $U_{\uparrow\downarrow}=0.1$ and $U_{\uparrow\downarrow}=2.0$, respectively. The rung current correlator also changes from an algebraic decay at $U_{\uparrow\downarrow}=0.1$ [Fig.~\ref{fig:12+12B2} (c)] to an exponential decay at $U_{\uparrow\downarrow}=2.0$ [Fig.~\ref{fig:12+12B2} (d)]. In contrast to the previous two cases, the chiral currents gradually decrease to zero and then change their directions. The scaling of the von Neumann EE versus the subsystem sizes at large $U_{\uparrow\downarrow}$ suggests that the spin density wave is gapped [Fig.~\ref{fig:12+12B3} (a)]. The derivative of the ground state energy with respect to $U_{\uparrow\downarrow}$ also implies that the phase transition is continuous [Fig.~\ref{fig:12+12B3} (b)]. It is also possible to analyze the large $U_{\uparrow\downarrow}$ limit using perturbation theory, but this leads to a very complicated spin model from which no useful information can be obtained since it has never been investigated before. 

\subsection{Meissner superfluid and the same magnetic field}

For one-component hard-core bosons with $\nu=1/4$ and $\phi=2\pi/5$, it has been shown that the ground state is a gapless Meissner superfluid~\cite{Greschner2015,Greschner2016}. This phase features quasi-long-range superfluid order, a chiral current, and central charge $1$. The two-component system has the same ${\mathbb Z}_{2}$ symmetry as in the first subsection. Numerical results suggest that there is one phase transition as $U_{\uparrow\downarrow}$ increases. The state at large $U_{\uparrow\downarrow}$ is a gapped vortex phase that breaks the ${\mathbb Z}_{2}$ symmetry.

\begin{figure}[ht]
\centering
\includegraphics[width=0.48\textwidth]{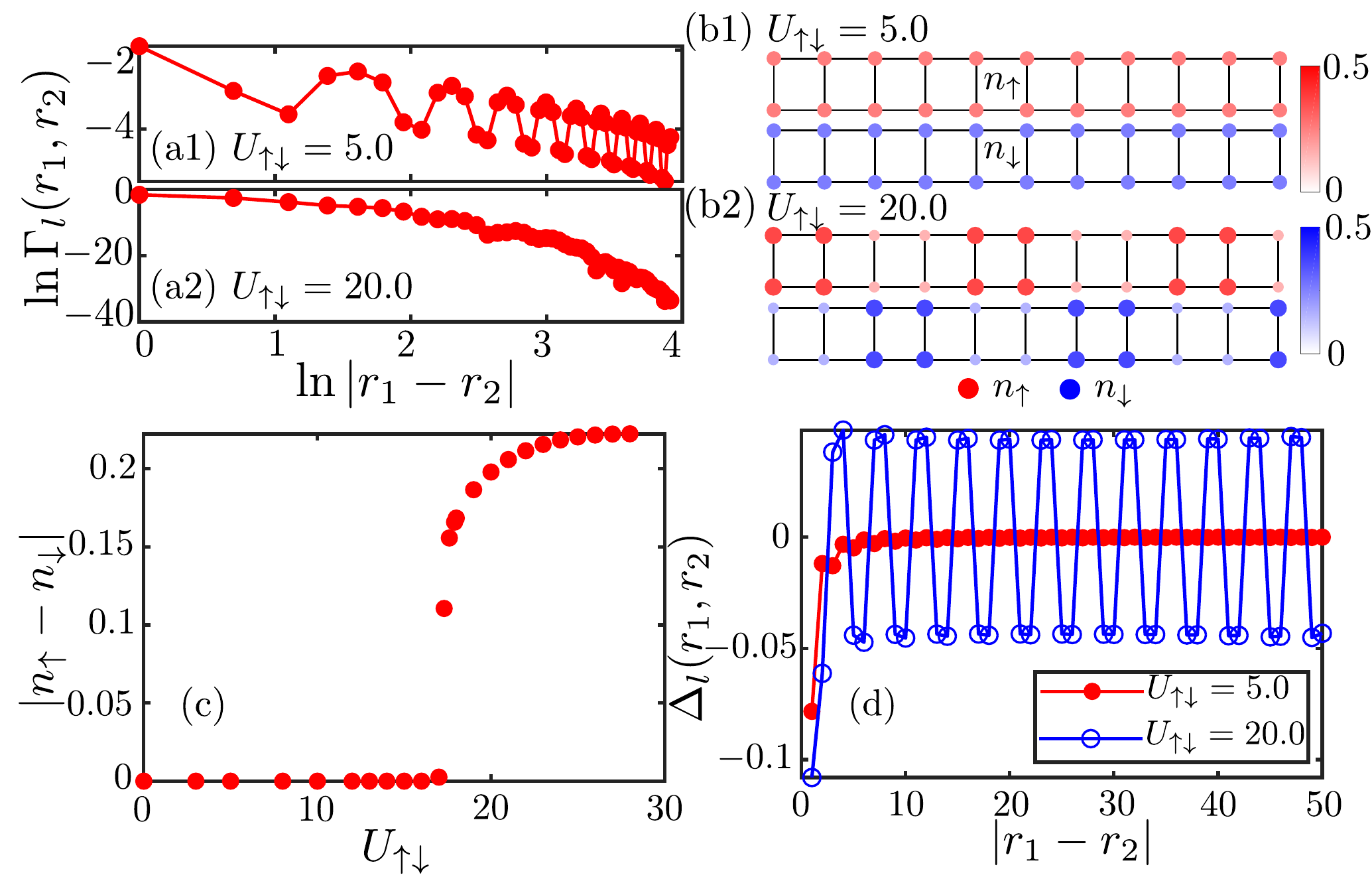}
\caption{Numerical results at filling factor $1/4+1/4$ where the two types of bosons have the same magnetic field. (a) The charge correlator $\Gamma_{l}(r_{1},r_{2})$ on the first leg. (b1-b2) The particle density profiles. In each subpanel, the top (bottom) one is for spin-up (spin-down) bosons. (c) The absolute difference of particle numbers on the first leg. (d) The density difference correlator $\Delta_{l}(r_{1},r_{2})$ on the first leg.}
\label{fig:14+14A1}
\end{figure}

The charge correlator $\Gamma_{\sigma,l}(r_{1},r_{2})$ decays algebraically when $U_{\uparrow\downarrow} \lesssim 18$. For the $U_{\uparrow\downarrow}=5.0$ case in Fig.~\ref{fig:14+14A1} (a), the curve can be fitted using Eq.~\ref{eq:SuperFluidOrder} with $q=0.627$. In constract, $\Gamma_{\sigma,l}(r_{1},r_{2})$ displays an exponential decay when $U_{\uparrow\downarrow}$ is sufficiently large. This implies that the superfluid order is destroyed by a phase transition. This transition is also accompanied by the emergence of density modulation as shown in Figs.~\ref{fig:14+14A1} (b), which persists up to the largest $U_{\uparrow\downarrow}$ that we have checked. The absolute difference of particle numbers on the first leg is shown in Fig.~\ref{fig:14+14A1} (c). The ${\mathbb Z}_{2}$ symmetry breaking at $U_{\uparrow\downarrow}=20$ is manifest from the density difference correlator $\Delta_{l}(r_{1},r_{2})$ in Fig.~\ref{fig:14+14A1} (d).

\begin{figure}[ht]
\centering
\includegraphics[width=0.48\textwidth]{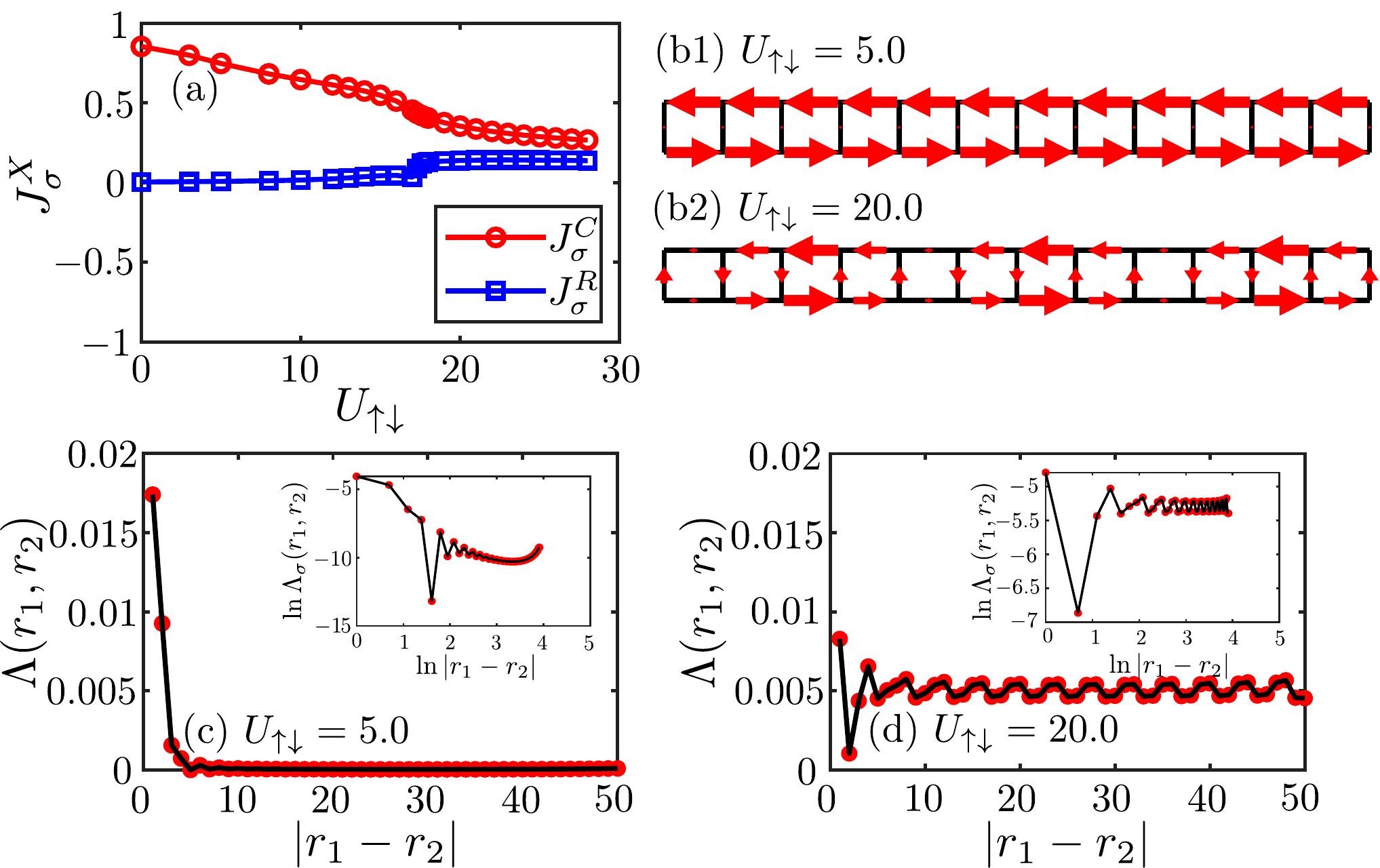}
\caption{Numerical results at filling factor $1/4+1/4$ where the two types of bosons have the same magnetic field. (a) The chiral current and the average rung current of the spin-up bosons. (b1-b2) The current pattern of the spin-up bosons. (c-d) The rung current correlator $\Lambda_{\sigma}(r_{1},r_{2})$ of the spin-up bosons.}
\label{fig:14+14A2}
\end{figure}

\begin{figure}[ht]
\centering
\includegraphics[width=0.48\textwidth]{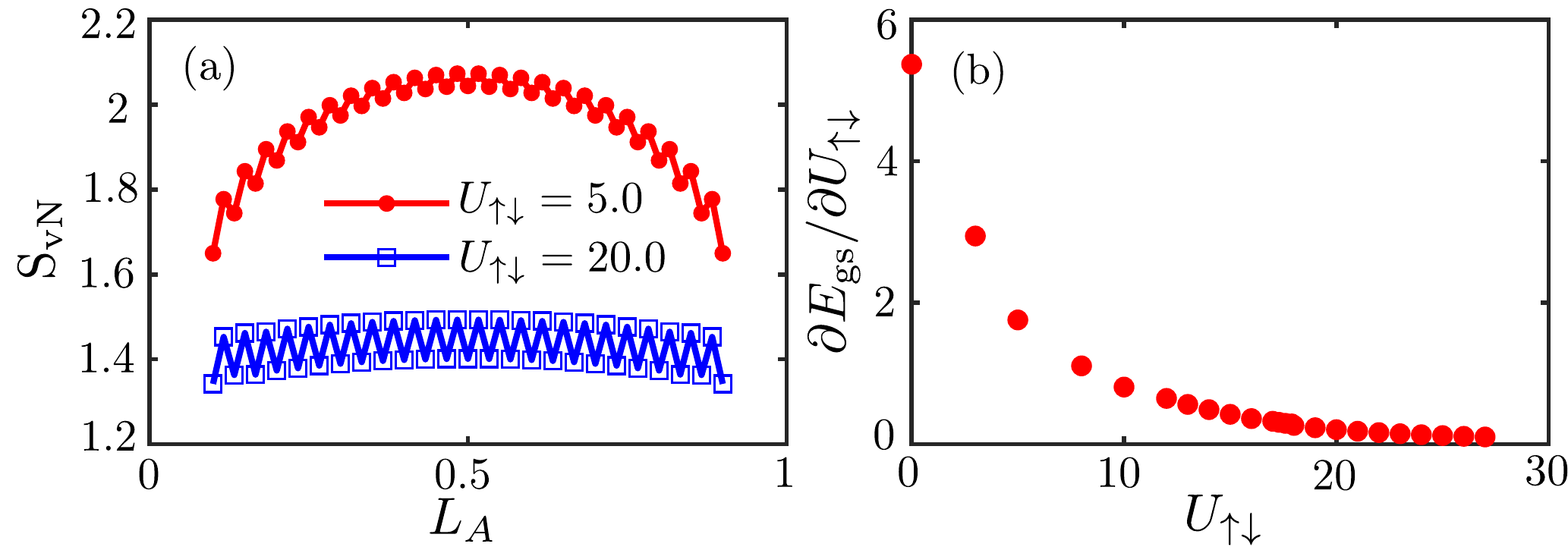}
\caption{Numerical results at filling factor $1/4+1/4$ where the two types of bosons have the same magnetic field. (a) The von Neumann EE versus subsystem size. The fitting parameters at $U_{\uparrow\downarrow}=5.0$ using Eq.~(\ref{eq:Cardy}) are $c=0.828$ and $g=1.151$. (b) The derivative of the ground state energy with respect to $U_{\uparrow\downarrow}$.}
\label{fig:14+14A3}
\end{figure}

The chiral current $J^{C}_{\sigma}$ and average rung current $J^{R}_{\sigma}$ are presented in Fig.~\ref{fig:14+14A2}. A nonzero $J^{R}_{\sigma}$ appears when the system passes the phase transition. Two current patterns at $U_{\uparrow\downarrow}=5.0$ and $U_{\uparrow\downarrow}=20.0$ are shown in Fig.~\ref{fig:14+14A2} (b), where one can clearly see chiral currents and vortex structures, respectively. The rung current correlator $\Lambda_{\sigma}(r_{1},r_{2})$ exhibits an exponential decay at $U_{\uparrow\downarrow}=5.0$ [Fig.~\ref{fig:14+14A2} (c)] but features long-range order at $U_{\uparrow\downarrow}=20$ [Fig.~\ref{fig:14+14A2} (d)]. These behaviors indicate that the system transits from a Meissner phase to a vortex phase as $U_{\uparrow\downarrow}$ increases. The scaling of the von Neumann EE versus the subsystem sizes at $U_{\uparrow\downarrow}{\gtrsim}18$ suggests that the vortex phase is gapped [Fig.~\ref{fig:14+14A3} (a)]. The derivative of the ground-state energy with respect to $U_{\uparrow\downarrow}$ also implies that the phase transition is continuous [Fig.~\ref{fig:14+14A3} (b)].

\subsection{Meissner superfluid and opposite magnetic fields}

As we have seen before, changing the fluxes for the two components to be opposite altered the behavior at $\nu_{\uparrow}=\nu_{\downarrow}=1/2$. This also turns out to be the case at $\nu_{\uparrow}=\nu_{\downarrow}=1/4$. In fact, we find that there is no phase transition if the two components have opposite magnetic fields. The quasi-long-range superfluid order gets weaker, but the density profile remains uniform up to $U_{\uparrow\downarrow}=50$. The amplitude of the chiral current decreases but the average rung current remains zero and there is no long-range correlation in the rung current.

\section{Conclusion}
\label{conclude}

In summary, we have studied quantum phases of two-component bosons on two-leg Harper-Hofstadter ladders at filling factors $1/2+1/2$ and $1/4+1/4$. The two types of bosons have either the same or opposite magnetic fields. The properties of the system are investigated by computing charge correlator, density profile, density difference correlator, particle currents, particle current correlators, entanglement entropy, and energy derivative. For the $1/2+1/2$ filling in both scenarios, the system transits to a gapped Meissner phase with spin density wave order as $U_{\uparrow\downarrow}$ increases. This state disappears at very large $U_{\uparrow\downarrow}$ if the two components have the same magnetic fields. For the $1/4+1/4$ filling, there is a phase transition to a gapped vortex phase if the two components have the same magnetic fields but no phase transition is observed if the two components have opposite magnetic fields. This work suggests that adding an internal degree of freedom to particles on the Harper-Hofstadter lattice can produce interesting results. We hope many other phenomena in multi-component system would be revealed in future works. 

\section*{Acknowledgment}

This work was supported by National Key R\&D Program of China under Grant No. 2018YFA0307200, the NNSF of China under grant No. 11574100, Key R\&D Program of Guangdong province under Grant No. 2019B030330001 (JDC and ZFX), the DFG through project A06 (HHT) of SFB 1143 (project-id 247310070), and the NNSF of China under grant No. 11804107 and startup grant of HUST (YHW).


\begin{widetext}

\section*{Appendix: Details about the effective spin model}
\label{Appendix}

This appendix explains in detail how to derive the effective spin-1/2 model in Eq.~(\ref{eq:spin}) of the main text. The hopping term and the interaction term in Eq.~(\ref{eq:model}) are denoted as $H_{0}$ and $H_{\rm int}$. The projection operator into the singly-occupied subspace is defined as $P_{0}$ and its complement is $P_{1}=1-P_{0}$. The two basis states of the spin-1/2 model are $|\uparrow\rangle$ and $|\downarrow\rangle$. The effective model can be constructed by computing the matrix elements for two neighboring sites along the leg direction and the rung direction. In both cases, the two sites are labeled as ${\rm I}$ and ${\rm II}$. The hopping strength is denoted as $T$, which would be $t_{x}$ for two sites on the same leg and $t_{y}e^{-ir\phi}$ if they are on the same rung. The matrix elements in second-order perturbation theory are
\begin{eqnarray}
\langle a|H^{(2)}_{\rm eff}| b \rangle &=& \langle a| H_{0} P_{1} \frac{1}{E^{(0)}-H_{\rm int}} P_{1} H_{0}|b\rangle \nonumber \\
&=& \sum_{\sigma=\uparrow,\downarrow} \langle \uparrow ,\uparrow | ( T a^{\dag}_{\sigma,{\rm I}} a_{\sigma,{\rm II}} + T^{*} a^{\dag}_{\sigma,{\rm II}} a_{\sigma,{\rm I}} ) P_{1} \frac{1}{E^{(0)}-H_{\rm int}} P_{1} (T a^{\dag}_{\sigma,{\rm I}} a_{\sigma,{\rm II}} + T^{*} a^{\dag}_{\sigma,{\rm II}} a_{\sigma,{\rm I}}) |\uparrow,\uparrow \rangle 
\end{eqnarray}
where $|a\rangle$ and $|b\rangle$ is one of the four states $|\uparrow,\uparrow\rangle,|\uparrow,\downarrow\rangle,|\downarrow,\uparrow\rangle,|\downarrow,\downarrow\rangle$ and the zeroth order energy $E^{(0)}$ is actually zero. Explicit calculations result in
\begin{eqnarray}
&& \langle \uparrow,\uparrow |H^{(2)}_{\rm eff}|\uparrow,\uparrow \rangle = -\frac{4|T|^{2}}{U_{0}}, \quad \langle \uparrow ,\downarrow |H^{(2)}_{\rm eff}|\uparrow ,\downarrow\rangle = -\frac{2|T|^{2}}{U_{\uparrow\downarrow}}, \nonumber \\
&& \langle \uparrow ,\downarrow |H^{(2)}_{\rm eff}| \downarrow,\uparrow\rangle = -\frac{2|T|^{2}}{U_{\uparrow\downarrow}}, \quad \langle \downarrow,\downarrow |H^{(2)}_{\rm eff}|\downarrow,\downarrow \rangle = -\frac{4|T|^{2}}{U_{0}}.
\end{eqnarray}

The two-site effective Hamiltonian can be written as
\begin{eqnarray}
H^{(2)}_{\rm eff} &=&
\begin{pmatrix}
-\frac{4|T|^{2}}{U_{0}} &  &  &  \\
& -\frac{2|T|^{2}}{U_{\uparrow\downarrow}} & -\frac{2|T|^{2}}{U_{\uparrow\downarrow}} & \\
& -\frac{2|T|^{2}}{U_{\uparrow\downarrow}} & -\frac{2|T|^{2}}{U_{\uparrow\downarrow}} & \\
&  &  & -\frac{4|T|^{2}}{U_{0}}
\end{pmatrix} \nonumber \\
&=& -|T|^{2} \frac{1}{U_{\uparrow\downarrow}} (\sigma^{x}\otimes\sigma^{x}+\sigma^{y}\otimes\sigma^{y}) -2|T|^{2} \left( \frac{1}{U_{0}}-\frac{1}{2U_{\uparrow\downarrow}} \right) \sigma^{z}\otimes\sigma^{z} -2|T|^{2} \left( \frac{1}{U_{0}}+\frac{1}{2U_{\uparrow\downarrow}} \right) \sigma^{0}\otimes\sigma^{0}
\label{eq:TwoSiteHeff}
\end{eqnarray}
in the basis $\{|\uparrow,\uparrow\rangle ,|\uparrow,\downarrow \rangle,|\downarrow,\uparrow\rangle,|\downarrow,\downarrow\rangle\}$. The spin-spin interaction along the leg is
\begin{eqnarray}
H_{\rm leg} = -J \sum_{l=1,2} \sum_{r} (S^{x}_{l,r} S^{x}_{l,r} + S^{y}_{l,r} S^{y}_{l,r} ) + J \sum_{l=1,2} \sum_{r} S^{z}_{l,r} S^{z}_{l,r},
\end{eqnarray}
and that along the rung is
\begin{eqnarray}
H_{\rm rung} = -J \sum_{r} (S^{x}_{1,r} S^{x}_{2,r} + S^{y}_{1,r} S^{y}_{2,r} ) + J \sum_{r} S^{z}_{1,r} S^{z}_{2,r}
\end{eqnarray}
with $J=4|T|^{2}/U_{\uparrow\downarrow}$. The effective Hamiltonian in Eq.~(\ref{eq:spin}) is obtained after a sublattice rotation
\begin{eqnarray}
&& \text{leg 1}: \quad S^{x}_{l=1,2j-1} \rightarrow -S^{x}_{l=1,2j-1}, \quad S^{y}_{l=1,2j-1} \rightarrow -S^{y}_{l=1,2j-1}, \quad S^{z}_{l=1,2j-1} \rightarrow S^{z}_{l=1,2j-1}; \nonumber \\
&& \text{leg 2}: \quad S^{x}_{l=2,2j} \rightarrow -S^{x}_{l=2,2j}, \quad S^{y}_{l=2,2j} \rightarrow -S^{y}_{l=2,2j}, \quad S^{z}_{l=2,2j} \rightarrow S^{z}_{l=2,2j}
\end{eqnarray}

\end{widetext}


\bibliography{ReferCollect}

\end{document}